\documentclass[preprint,aps,showpacs,showkeys]{revtex4}

\usepackage[pdftex]{graphicx}
\usepackage{amsmath,amsfonts,amssymb,amsxtra}

\usepackage[final,colorlinks=true,linkcolor=red,citecolor=blue]{hyperref}
\begin{document}
\title{Relations between Clifford 
algebra and Dirac matrices in the presence of families
} %
%
\author{D. Lukman${}^2$, M. Komendyak${}^3$, N.S. Manko\v c Bor\v stnik${}^1$\\ 
\small{${}^1$University of Ljubljana, Department of Physics, 
SLO, ${}^2$Current address: University of Maribor, CAMTP, SLO,
${}^3$University of Warwick, Department of Physics,
CV4 7AL, UK}\\
}
%
\begin{abstract}
The internal  degrees of freedom of fermions are in the {\it spin-charge-family}
theory~\cite{norma93,n2014matterantimatter,IARD2016,normaJMP2015,nh02,nh03} 
described by the Clifford algebra objects, which are superposition of an odd number of 
$\gamma^a$'s. Arranged into irreducible representations of "eigenvectors" of the 
Cartan subalgebra of the Lorentz algebra $S^{ab}$ 
$(= \frac{i}{2} \gamma^a \gamma^b|_{a \ne b})$ these objects form $2^{\frac{d}{2}-1}$
families with $2^{\frac{d}{2}-1}$ family members each. Family members of each family
offer the description of all the observed quarks and leptons and antiquarks and antileptons, 
appearing in families. Families are reachable by $\tilde{S}^{ab}$ 
$=\frac{1}{2} \tilde{\gamma}^a \tilde{\gamma}^b|_{a \ne b}$. 
Creation operators, carrying the family member and family quantum numbers form the
basic vectors. The action of the operators $\gamma^a$'s, $S^{ab}$, $\tilde{\gamma}^a$'s 
and $\tilde{S}^{ab}$, applying on the basic vectors, manifests as  matrices.
%
In this paper the basic vectors in $d=(3+1)$ Clifford space are discussed, chosen in a way
that the matrix representations of $\gamma^a$ and of 
$S^{ab}$ 
coincide for each family quantum number, determined by $\tilde{S}^{ab} $, 
with the Dirac  matrices. 
The appearance of charges in Clifford space is discussed by embedding $d=(3+1)$ space
into $d=(5+1)$-dimensional space.

\end{abstract}





\maketitle

\section{Introduction}
\label{introduction}

In the Grassmann graded algebra of anticommuting coordinates $\theta^a$ there are  in 
$d$-dimensional space $2^d$ vectors, which define, together with the corresponding 
derivatives $\frac{\partial}{\partial \theta_a}$, two kinds of the Clifford algebra objects: 
$\gamma^a$  and $\tilde{\gamma}^a$~\cite{norma93,nh02,nh03,nh2018}, both with the
anticommutation properties of the Dirac $\gamma^a$ matrices, while the anticommutators  
among $\gamma^a$  and $\tilde{\gamma}^b$ are equal to zero. 
\begin{eqnarray}
\label{cliffcomrel}
\{\gamma^a, \gamma^b\}_{+} &=& 2\eta^{a b} =  \{\tilde{\gamma}^{a}, 
\tilde{\gamma}^{b}\}_{+}\,, \quad \{\gamma^{a}, \tilde{\gamma}^{b}\}_{+} = 0\,,
\nonumber\\
  (\gamma^{a})^{\dagger}& =& \eta^{aa}\, \gamma^{a}\,, \quad
(\tilde{\gamma}^{a})^{\dagger} =  \eta^{a a}\, \tilde{\gamma}^{a}\,, \nonumber\\
S^{ab} &=&\frac{i}{4}  (\gamma^a \gamma^b - \gamma^b \gamma^a)\,,\quad  
\tilde{S}^{ab} =\frac{i}{4}  (\tilde{\gamma}^a \tilde{\gamma}^b - \tilde{\gamma}^b
\tilde{\gamma}^a)\,,\nonumber\\
\{ S^{ab}, \tilde{S}^{ab}\}_{+} &=&0\,, \nonumber\\
(a,b)&=&(0,1,2,3,5,\cdots,d)\,.
\end{eqnarray}
The two Clifford algebras, $\gamma^a$'s  and $\tilde{\gamma}^a$'s, are obviously 
completely
independent and form two independent spaces, each with $2^d$ vectors~\cite{n2019PIPII}.

Sacrificing the space of $\tilde{\gamma}^a$'s by defining 
\begin{eqnarray}
\tilde{\gamma}^a B (\gamma^a) &=&(-)^B\, i \, B \gamma^a\,,
\label{tildegammareduced}
\end{eqnarray}
with $(-)^B = -1$, if $B$ is an odd product of $\gamma^a$'s, otherwise 
$(-)^B = 1$~\cite{nh03}, we end up with vector space of $2^{d}$ degrees of freedom, 
defined by $\gamma^a$'s only. 

A general vector can correspondingly be written as
\begin{eqnarray}
\label{grassmannfermion}
{\bf B} &=& \sum_{k=0}^{d}\, a_{a_{1} a_{2} \dots a_{k}}\,
\gamma^{a_1} \gamma^{a_2} \dots \gamma^{a_k} |\psi_{o}>\,,
\quad a_{i}\le a_{i+1}\,,
\end{eqnarray}
where $|\psi_{o}>$ is the vacuum state.

We arrange these vectors as products of nilpotents and projectors 
\begin{eqnarray}
\stackrel{ab}{(k)}:&=& 
\frac{1}{2}(\gamma^a + \frac{\eta^{aa}}{ik} \gamma^b)\,,\quad 
(\stackrel{ab}{(k)})^2=0\,.
\nonumber\\
\stackrel{ab}{[k]}:&=&
\frac{1}{2}(1+ \frac{i}{k} \gamma^a \gamma^b)\,, \quad (\stackrel{ab}{[k]})^2=
\stackrel{ab}{[k]}\,, 
\label{signature}
\end{eqnarray}
where $k^2 = \eta^{aa} \eta^{bb}$.  Their Hermitian conjugated values  follow from 
Eq.~(\ref{cliffcomrel}).  
%
\begin{eqnarray}
\stackrel{ab}{(k)}^{\dagger}=\eta^{aa}\stackrel{ab}{(-k)},\quad
\stackrel{ab}{[k]}^{\dagger}= \stackrel{ab}{[k]}\,.
\label{graphher}
\end{eqnarray}
Vectors in Clifford space  are chosen to be eigenstates of the Cartan subalgebra, 
Eq.~(\ref{cartan}), of the generators of the Lorentz transformations $S^{ab}$
in the internal space of $\gamma^a$'s  
\begin{eqnarray}
S^{03}, S^{12}, S^{56}, \cdots, S^{d-1 \;d}\,, \nonumber\\
\tilde{S}^{03}, \tilde{S}^{12}, \tilde{S}^{56}, \cdots,  \tilde{S}^{d-1\; d}\,,
\label{cartan}
\end{eqnarray}
with the eigenvalues 
%
$S^{ab}\stackrel{ab}{(k)}=\frac{1}{2}k \stackrel{ab}{(k)}$, 
$S^{ab}\stackrel{ab}{[k]}=\frac{1}{2}k \stackrel{ab}{[k]}$.
%
All the relations of Eq.~(\ref{cliffcomrel}) remain unchanged after the assumption of
Eq.~(\ref{grassmannfermion}), while 
each irreducible representation of the Lorentz algebra $S^{ab}$ receives the additional
quantum number $f$, defined by $\tilde{S}^{ab}$ 
\begin{eqnarray}
\label{grapheigen}
S^{ab} \,\stackrel{ab}{(k)} = \frac{k}{2}  \,\stackrel{ab}{(k)}\,,\quad && \quad
\tilde{S}^{ab}\,\stackrel{ab}{(k)} = \frac{k}{2}  \,\stackrel{ab}{(k)}\,,\nonumber\\
S^{ab}\,\stackrel{ab}{[k]} =  \frac{k}{2}  \,\stackrel{ab}{[k]}\,,\quad && \quad 
\tilde{S}^{ab} \,\stackrel{ab}{[k]} = - \frac{k}{2}  \,\,\stackrel{ab}{[k]}\,.
\end{eqnarray}
 Eq.~(\ref{grapheigen}) demonstrates that the eigenvalues of $S^{ab}$ on nilpotents and 
projectors generated by $\gamma^a$'s differ from the eigenvalues of $\tilde{S}^{ab}$. 
Nilpotents are the superposition of odd number of $\gamma^a$'s,  projectors have an even 
Clifford character.

States, which are products of projectors and nilpotents, have well defined handedness 
of both kinds, $\Gamma^{(d)}$ and $\tilde{\Gamma}^{(d)}$ 
\begin{eqnarray}
\Gamma^{(d)} :&=&(i)^{d/2}\; \;\;\;\;\;\prod_a \quad (\sqrt{\eta^{aa}} \gamma^a)\,, 
\quad {\rm if } \quad d =2n\,,\nonumber\\ 
\label{sabgamma}
%
%
\tilde{\Gamma}^{(d)} :&=&(i)^{d/2}\; \;\;\;\;\;\prod_a \quad (\sqrt{\eta^{aa}}
 \tilde{\gamma}^a)\,,
\quad {\rm if } \quad d =2n\,. 
\end{eqnarray}

The {\it spin-charge-family} theory~\cite{norma93,n2014matterantimatter,IARD2016,%
normaJMP2015,nh02,nh03} of N.S. Manko\v c Bor\v stnik uses  products of nilpotents,  
$\stackrel{ab}{(k)}=\frac{1}{2}(\gamma^a + \frac{\eta^{aa}}{ik} \gamma^b)$, and 
projectors, $\stackrel{ab}{[k]} =\frac{1}{2}(1 + \frac{i}{k} \gamma^a\gamma^b)$,
 to define $2^d$ vectors in this space of  the Clifford graded 
algebra~\cite{norma93,n2014matterantimatter,IARD2016,normaJMP2015}.  In this theory 
$S^{ab}$ determine in $d=(3+1)$ space, which  is a part of $d=(13+1)$-dimensional 
space, spins and charges of quarks and leptons, while $\tilde{S}^{ab}$ determine
families of quarks and leptons.

It is interesting to notice~(\cite{n2019PIPII,nh2018} and references therein): {\it Vectors, 
which are superposition of an odd products of $\gamma^a$'s, anticommute. 
Half of them can be taken as creation operators and the other half as annihilation operators.
These creation and annihilation operators then fulfill the anticommutation relations postulated 
by} Dirac~\cite{Dirac1928} {\it for second quantized fermions, and consequently 
explain}~\cite{n2019PIPII,nh2018} {\it the Dirac's postulate}.

In Sect.~\ref{properties} the properties of products of nilpotents and projectors are discussed,  
arranged into eigenvectors of the Cartan subalgebra,  when $d=(3+1)$-dimensional 
space is embedded into $d=(5+1)$-dimensional space. Nilpotents and projectors 
define the internal vector space of fermions so that the spin in $d= (5,6)$ 
manifests as a charge of fermions in $d=(3+1)$.  

In Sect.~\ref{basis} the matrix representation of vectors are presented.

\section{Properties of vectors in Clifford space}
\label{properties}

In Refs.~\cite{n2019PIPII,nh2018} the fact that the Clifford vectors, spanned by products 
of an odd number of $\gamma^a$'s, are fulfilling the anticommutation relations postulated 
by Dirac for the second quantized fermions, are discussed. Let us illustrate how this happens 
in the case that $d=(5+1)$. 

Let us denote  vectors in $d=(5+1)$ of an odd Clifford character (they are superposition of an 
odd products of $\gamma^a$'s), presented in Table~\ref{cliff basis5+1.} as  products of 
nilpotents and projectors, by $\hat{b}^{f \dagger}_{m}$ (the third column on 
Table~\ref{cliff basis5+1.}).  The member quantum number $ m=(ch,s)$ includes the 
charge $ch$  and the spin $s$, the charge concerns the eigenvalue of $S^{56}$, the spin
the eigenvalue of $S^{12}$. The corresponding Hermitian conjugated partner (the fourth 
column on Table~\ref{cliff basis5+1.})  is denoted by
$(\hat{b}^{f \dagger}_{m})^{\dagger}=\hat{b}^{f}_{m}$.

The first member $m=(\frac{1}{2}, \frac{1}{2})$ of the first family $a$, which is the product 
of three nilpotents, is correspondingly denoted by 
$\hat{b}^{a \dagger}_{(\frac{1}{2}, \frac{1}{2})}=$
$\stackrel{03}{(+i)}\,\stackrel{12}{(+)}| \stackrel{56}{(+)}$. All the other vectors 
$\hat{b}^{f \dagger}_{m}$ of the family $f=a$ follow by the application of $S^{ab}$. 
The families $\hat{b}^{f \dagger}_{m}$, $f=(b,c,d)$ follow from $f=a$ by the
application of $\tilde{S}^{ab}$. The Hermitian conjugated partners follow by the application of
Eq.~(\ref{cliffcomrel}). 

Table~\ref{cliff basis5+1.}, taken from Table IV of Ref.~\cite{nh2018}, represents four  
families of Clifford odd vectors and their Hermitian conjugated partners. All the families have 
the same quantum numbers  $m$ of the corresponding members, $(S^{03}, S^{12}, S^{56})$,
each family carries its own family quantum number $f$.
 \begin{table}
\begin{tiny}
 \begin{center}
 \begin{tabular}{|r|r|r|r|r r r r r r r |r|}
 \hline
$ f {\rm (amily)}\, m $&$(ch,s)$&$\hat{b}^{f \dagger}_{m}$&$\hat{b}^{f}_{m}$&
$S^{03}$&$ S^{1 2}$&$S^{5 6}$&$\Gamma^{(3+1)}$ &
$\tilde{S}^{03}$&$\tilde{S}^{1 2}$& $\tilde{S}^{5 6}$\\
\hline
$a\; 1$&$ (\frac{1}{2},\frac{1}{2})$&$
\stackrel{03}{(+i)}\,\stackrel{12}{(+)}| \stackrel{56}{(+)}$&
$
{\scriptstyle (-)} \stackrel{56}{(-)}|{\scriptstyle (-)} 
\stackrel{12}{(-)} \stackrel{03}{(-i)}$&$\frac{i}{2}$&$\frac{1}{2}$&$\frac{1}{2}$&$1$
&$\frac{i}{2}$&$\frac{1}{2}$&$\frac{1}{2}$\\
$a\;2$ &$(\frac{1}{2},-\frac{1}{2})$&$
\stackrel{03}{[-i]}\,\stackrel{12}{[-]}|\stackrel{56}{(+)}$&$
 {\scriptstyle (-)} \stackrel{56}{(-)}| \stackrel{12}{[-]} \stackrel{03}{[-i]}$&
$-\frac{i}{2}$&$-\frac{1}{2}$&$\frac{1}{2}$&$1$
&$\frac{i}{2}$&$\frac{1}{2}$&$\frac{1}{2}$\\

$a\;3$ &$(-\frac{1}{2},\frac{1}{2})$&$
 \stackrel{03}{[-i]}\,\stackrel{12}{(+)}|\stackrel{56}{[-]}$&$
 \stackrel{56}{[-]}|{\scriptstyle (-)} \stackrel{12}{(-)} \stackrel{03}{[-i]}$&
$-\frac{i}{2}$&$ \frac{1}{2}$&$-\frac{1}{2}$&$-1$
&$\frac{i}{2}$&$\frac{1}{2}$&$\frac{1}{2}$\\
$a\;4$ &$(-\frac{1}{2},-\frac{1}{2})$&$
\stackrel{03}{(+i)}\,\stackrel{12}{[-]}|\stackrel{56}{[-]}$&$
\stackrel{56}{[-]}| \stackrel{12}{[-]} \stackrel{03}{(-i)}$&
$\frac{i}{2}$&$- \frac{1}{2}$&$-\frac{1}{2}$&$-1$
&$\frac{i}{2}$&$\frac{1}{2}$&$\frac{1}{2}$\\
\hline 
$b\;1$&$(\frac{1}{2},\frac{1}{2})$&$
\stackrel{03}{[+i]}\,\stackrel{12}{[+]}| \stackrel{56}{(+)}$&
$
{\scriptstyle (-)} \stackrel{56}{(-)}|\stackrel{12}{[+]} \stackrel{03}{[+i]}$&$\frac{i}{2}$&
$\frac{1}{2}$&$\frac{1}{2}$&$1$&$-\frac{i}{2}$&$-\frac{1}{2}$&$\frac{1}{2}$\\
$b\;2$ &$(\frac{1}{2},-\frac{1}{2})$&$
\stackrel{03}{(-i)}\,\stackrel{12}{(-)}|\stackrel{56}{(+)}$&$
 {\scriptstyle (-)} \stackrel{56}{(-)}| {\scriptstyle (-)} \stackrel{12}{(+)}
 \stackrel{03}{(+i)}$&
$-\frac{i}{2}$&$-\frac{1}{2}$&$\frac{1}{2}$&$1$
&$-\frac{i}{2}$&$-\frac{1}{2}$&$\frac{1}{2}$\\
$b\;3$ &$(-\frac{1}{2},\frac{1}{2})$&$
\stackrel{03}{(-i)}\,\stackrel{12}{[+]}|\stackrel{56}{[-]}$&$
 \stackrel{56}{[-]}| \stackrel{12}{[+]} \stackrel{03}{(+i)}$&
$-\frac{i}{2}$&$ \frac{1}{2}$&$-\frac{1}{2}$&$-1$
&$-\frac{i}{2}$&$-\frac{1}{2}$&$\frac{1}{2}$\\
$b\;4$ &$(-\frac{1}{2},-\frac{1}{2})$&$
\stackrel{03}{[+i]}\, \stackrel{12}{(-)}|\stackrel{56}{[-]}$&$
\stackrel{56}{[-]}|{\scriptstyle (-)}  \stackrel{12}{(+)} \stackrel{03}{[+i]}$&
$\frac{i}{2}$&$- \frac{1}{2}$&$-\frac{1}{2}$&$-1$
&$-\frac{i}{2}$&$-\frac{1}{2}$&$\frac{1}{2}$\\ 
%
%
 \hline
$c\;1$&$(\frac{1}{2},\frac{1}{2})$&$
\stackrel{03}{[+i]}\,\stackrel{12}{(+)}| \stackrel{56}{[+]}$&
$
 \stackrel{56}{[+]}|{\scriptstyle (-)}\stackrel{12}{(-)} \stackrel{03}{[+i]}$&$\frac{i}{2}$&
$\frac{1}{2}$&$\frac{1}{2}$&$1$&$-\frac{i}{2}$&$\frac{1}{2}$&$-\frac{1}{2}$\\
$c\;2$ &$(\frac{1}{2},-\frac{1}{2})$&$
\stackrel{03}{(-i)}\,\stackrel{12}{[-]}|\stackrel{56}{[+]}$&$
 \stackrel{56}{[+]}|  \stackrel{12}{[-]} \stackrel{03}{(+i)}$&
$-\frac{i}{2}$&$-\frac{1}{2}$&$\frac{1}{2}$&$1$
&$-\frac{i}{2}$&$\frac{1}{2}$&$-\frac{1}{2}$\\
$c\;3$ &$(-\frac{1}{2},\frac{1}{2})$&$
\stackrel{03}{(-i)}\,\stackrel{12}{(+)}|\stackrel{56}{(-)}$&$
 {\scriptstyle (-)}\stackrel{56}{(+)}| {\scriptstyle (-)} \stackrel{12}{(-)} \stackrel{03}{(+i)}$&
$-\frac{i}{2}$&$ \frac{1}{2}$&$-\frac{1}{2}$&$-1$
&$-\frac{i}{2}$&$\frac{1}{2}$&$-\frac{1}{2}$\\
$c\;4$ &$(-\frac{1}{2},-\frac{1}{2})$&$
\stackrel{03}{[+i]} \stackrel{12}{[-]}|\stackrel{56}{(-)}$&$
{\scriptstyle (-)} \stackrel{56}{(+)}|  \stackrel{12}{[-]} \stackrel{03}{[+i]}$&
$\frac{i}{2}$&$- \frac{1}{2}$&$-\frac{1}{2}$&$-1$
&$-\frac{i}{2}$&$\frac{1}{2}$&$-\frac{1}{2}$\\
\hline
$d\;1$&$(\frac{1}{2},\frac{1}{2})$&$
\stackrel{03}{(+i)}\,\stackrel{12}{[+]}| \stackrel{56}{[+]}$&
$
 \stackrel{56}{[+]}|\stackrel{12}{[+]} \stackrel{03}{(-i)}$&$\frac{i}{2}$&
$\frac{1}{2}$&$\frac{1}{2}$&$1$&$\frac{i}{2}$&$-\frac{1}{2}$&$-\frac{1}{2}$\\
$d\;2$ &$(\frac{1}{2},-\frac{1}{2})$&$
\stackrel{03}{[-i]}\,\stackrel{12}{(-)}|\stackrel{56}{[+]}$&$
 \stackrel{56}{[+]}| {\scriptstyle (-)} \stackrel{12}{(+)}
 \stackrel{03}{[-]}$&
$-\frac{i}{2}$&$-\frac{1}{2}$&$\frac{1}{2}$&$1$
&$\frac{i}{2}$&$-\frac{1}{2}$&$-\frac{1}{2}$\\
$d\;3$ &$(-\frac{1}{2},\frac{1}{2})$&$
\stackrel{03}{[-i]}\,\stackrel{12}{[+]}|\stackrel{56}{(-)}$&$
{\scriptstyle (-)} \stackrel{56}{(+)}| \stackrel{12}{[+]} \stackrel{03}{[-i]}$&
$-\frac{i}{2}$&$ \frac{1}{2}$&$-\frac{1}{2}$&$-1$
&$\frac{i}{2}$&$-\frac{1}{2}$&$-\frac{1}{2}$\\
$d\;4$ &$(-\frac{1}{2},-\frac{1}{2})$&$
\stackrel{03}{(+i)}\, \stackrel{12}{(-)}|\stackrel{56}{(-)}$&$
{\scriptstyle (-)} \stackrel{56}{(+)}|{\scriptstyle (-)}  \stackrel{12}{(+)} \stackrel{03}{(-i)}$&
$\frac{i}{2}$&$- \frac{1}{2}$&$-\frac{1}{2}$&$-1$
&$\frac{i}{2}$&$-\frac{1}{2}$&$-\frac{1}{2}$\\ 
\hline 
 \end{tabular}
 \end{center}
\end{tiny}
 \caption{\label{cliff basis5+1.}  The basic creation operators, which are sums of odd products
of $\gamma^a$'s, $\hat{b}^{f \dagger}_{m}$ and their annihilation partners $\hat{b}^{f}_{m}$
are presented for the $d= (5+1)$-dimensional case. Here $m=(ch, s)$,  $ch$ represents 
the spin in $d=(5,6)$, manifesting in $d=(3+1)$  the charge, and $s$ represents the spin, 
that is the eigenvalue of $S^{12}$, according to the choice of the Cartan subalgebra, 
Eq.~(\ref{cartan}).
The basic creation operators are the products of nilpotents and projectors, which are the 
"eigenstates" of the Cartan subalgebra generators, ($S^{03}$, $S^{12}$, $S^{56}$) and  
($\tilde{S}^{03}$, $\tilde{S}^{12}$, $\tilde{S}^{56}$), 
presented in Eq.~(\ref{cartan}). 
}
 \end{table}

Half of vectors, the eigenvectors of the Cartan subalgebra, Eq.~(\ref{cartan}), which are 
products of nilpotents and projectors, are odd products of $\gamma^a$'s and half of them
are even products of $\gamma^a$'s. On Table~\ref{cliff basis5+1.} only Clifford odd vectors 
are presented.

Let  us make a choice of the vacuum state~\cite{nh02,nh03,nh2018,n2019PIPII}. 
(In the case of a general even $d$ the normalization factor is 
$\frac{1}{\sqrt{2^{\frac{d}{2}-1}}}$, since the vacuum states, generated by projectors only,
follows from the starting products of $\frac{d}{2}$ projectors, let say $ \stackrel{03}{[-i]}\,
 \stackrel{12}{[-]}|\stackrel{56}{[-]}\,\dots\, \stackrel{d-1\,d}{[-]})$, by transforming all 
possible pairs of $[-]...[-]$, with $[-i]$ included, to $[+]...[+]$, creating therefore  
$2^{\frac{d}{2}-1}$ summands
\begin{eqnarray}  
|\psi_o>&=& (\frac{1}{\sqrt{2}})^2 \,(
\stackrel{03}{[-i]}\, \stackrel{12}{[-]}|\stackrel{56}{[-]} + 
\stackrel{03}{[+i]}\, \stackrel{12}{[+]}|\stackrel{56}{[-]} + 
\stackrel{03}{[+i]}\, \stackrel{12}{[-]}| \stackrel{56}{[+]} + 
\stackrel{03}{[-i ]}\, \stackrel{12}{[+]}|\stackrel{56}{[+]})\,|1>\,.
\label{vac}
\end{eqnarray} 
The reader can check, taking into account Eqs.~(\ref{cliffcomrel}, 
\ref{snmb:gammatildegamma}) or Eq.~(\ref{graphbinoms})
(taken from Ref.~\cite{IARD2016})
\begin{eqnarray}
\stackrel{ab}{[k]}\stackrel{ab}{[k]} &=& \stackrel{ab}{[k]}\,, 
\stackrel{ab}{[k]}\stackrel{ab}{[-k]}= 0\,, \;\;
\stackrel{ab}{(k)}\stackrel{ab}{[k]}= 0\,,
  \stackrel{ab}{(k)}\stackrel{ab}{[-k]} =  \stackrel{ab}{(k)}\,,
\label{graphbinoms}
\end{eqnarray}
%
that 
\begin{eqnarray} 
\hat{b}^{f \dagger}_{m} \,|\psi_o>:&=& |\psi^{f}_{m}>\,,\nonumber\\
\hat{b}^{f}_{m} \,|\psi_o>&=& 0 \cdot \,|\psi_o>\,,\nonumber\\
\{ \hat{b}^{f \dagger}_{m}\,, \hat{b}^{f'}_{m'}\}_{+} \,|\psi_o> &=&
\delta^{f f'}\delta_{m m'}\, |\psi_{o}>  \, ,\nonumber\\
\{ \hat{b}^{f \dagger}_{m}\,, \hat{b}^{f'\dagger}_{m'}\}_{+} &=& 0 \cdot \, |\psi_{o}> \,,
\nonumber\\
\{ \hat{b}^{f}_{m}\,, \hat{b}^{f'}_{m'}\}_{+} &=& 0\cdot \, |\psi_{o}> \,,
\nonumber\\
\forall \, m \, {\rm and }\, \forall \,f\,.
\label{creationvac}
\end{eqnarray} 
The relations among creation and annihilation operators in Eq.~(\ref{creationvac}) fulfill
all the Dirac's requirements for the second quantized fermions.

\subsection{Action}
\label{action}

The Lorentz invariant action for a free massless fermion, describing 
internal degrees of freedom in Clifford space,  is well known 
\begin{eqnarray}
{\cal A}\,  &=& \int \; d^dx \; \frac{1}{2}\, (\psi^{\dagger}\gamma^0 \, \gamma^a p_{a} \psi) +
 h.c.\,, 
\label{actionWeyl}
\end{eqnarray}
$p_{a} = i\, \frac{\partial}{\partial x^a}$. It leads to the Weyl equations of motion 
\begin{eqnarray}
\label{Weyl}
\gamma^a p_{a}  |\psi>&= & 0\,, 
\end{eqnarray}
which fulfill also the Klein-Gordon equation
\begin{eqnarray}
\label{LtoKG}
\gamma^a p_{a} \gamma^b p_b |\psi>&= &   
p^a p_a |\psi>=0\,,\nonumber\\
\end{eqnarray}
$\gamma^0$ appears in the 
action to take care of the Lorentz invariance of the action.

Solutions of equations of motion, Eq.~(\ref{Weyl}), for a free massless  fermion with 
momentum $p^{a}= (|p^0|, p^{1}, p^2, p^3, 0, 0)$  and  a particular charge  
$ \frac{1}{2}$, are for any family $f$ superposition of  basic vectors $ |\psi^{m}_{f}>= 
\hat{b}^{f \dagger}_{m}\, |\psi_o>$ with spin $\frac{1}{2}$  
 and  spin $-\frac{1}{2}$, both   multiplied by $e^{-i (p^0 x^0-\vec{p} \vec{x})}$, 
 (see Eq.~ (97) in  Ref.~\cite{nh2018}). 
Coefficients in the superposition depend on the momentum $p^{a}$.

\subsection{Creation and annihilation operators in $d=(3+1)$ space embedded in $d=(5+1)$
space}
\label{3+15+1}

The creation and annihilation operators of Table~\ref{cliff basis5+1.} are all of an odd 
Clifford character (they are superposition of odd products of $\gamma^a$'s). The rest of 
the $2^4$ creation operators of an even Clifford character can be found in 
Refs.~\cite{n2019PIPII,nh2018}. 

Taking into account Eq.~(\ref{cliffcomrel}) one recognizes  that $\gamma^a$'s 
transform  $\stackrel{ab}{(k)}$ into  $\stackrel{ab}{[-k]}$, never to $\stackrel{ab}{[k]}$, 
while $\tilde{\gamma}^a$'s transform  $\stackrel{ab}{(k)}$ into $\stackrel{ab}{[k]}$, never to 
$\stackrel{ab}{[-k]}$ 
\begin{eqnarray}
&&\gamma^a \stackrel{ab}{(k)}= \eta^{aa}\stackrel{ab}{[-k]},\; 
\gamma^b \stackrel{ab}{(k)}= -ik \stackrel{ab}{[-k]}, \; 
\gamma^a \stackrel{ab}{[k]}= \stackrel{ab}{(-k)},\; 
\gamma^b \stackrel{ab}{[k]}= -ik \eta^{aa} \stackrel{ab}{(-k)}\,,\nonumber\\
&&\tilde{\gamma^a} \stackrel{ab}{(k)} = - i\eta^{aa}\stackrel{ab}{[k]},\;
\tilde{\gamma^b} \stackrel{ab}{(k)} =  - k \stackrel{ab}{[k]}, \;
\tilde{\gamma^a} \stackrel{ab}{[k]} =  \;\;i\stackrel{ab}{(k)},\; 
\tilde{\gamma^b} \stackrel{ab}{[k]} =  -k \eta^{aa} \stackrel{ab}{(k)}\,. 
\label{snmb:gammatildegamma}
\end{eqnarray}
With the knowledge presented in Eq.~(\ref{snmb:gammatildegamma}) it is not difficult to
reproduce Table~\ref{tableBasVecd4m},  representing vectors that belong to $d=(3+1)$
space. Vectors carry no charge and have either an odd or an even Clifford character. 
Multiplying these vectors by  the appropriate charge (that is by either the  nilpotent,  if the 
$d=(3+1)$ part has an even Clifford character, or the projector, if the $d=(3+1)$ part has 
an odd Clifford character --- both must be the eigenfunction of $S^{56}$) we end up with 
the Clifford odd vectors from Table~\ref{cliff basis5+1.}.

The properties of vectors of Table~\ref{tableBasVecd4m} are analyzed in details in order that
the correspondence with the Dirac $\gamma$ matrices in $d=(3+1)$ space is easy to
recognize. Superposition of vectors with the spin $\pm \frac{1}{2}$ (either Clifford even 
or odd) solve the equations of motion, Eq.~(\ref{Weyl}), for free massless fermions.
\begin{table}
\begin{tiny}
\begin{center}
  \begin{tabular}{|c| c |r r r r r r r r |r r r r r r |}
    \hline 
    & $\psi^f_m$ & $\gamma_0\,\psi^f_m$ & 
$\gamma_1\,\psi^f_m$ & $\gamma_2\,\psi^f_m$     &$\gamma_3\,\psi^f_m$ &
$\Tilde{\gamma}_0\,\psi^f_m$ &$\Tilde{\gamma}_1\,\psi^f_m$ & 
$\Tilde{\gamma}_2\,\psi^f_m$  &$\Tilde{\gamma}_3\,\psi^f_m$ & 
$S^{03}$ & $S^{12}$ &$\Tilde{S}^{03}$ &
$\Tilde{S}^{12}$ & $\Gamma^{(3+1)}$ & \vphantom{$b\biggm|b$} 
$\Tilde{\Gamma}^{(3+1)}$\\
    \hline
    $\psi^a_1$ & $(+i) (+)$ & $\psi^a_3$ & $\psi^a_4$ & $i\psi^a_4$ & $\psi^a_3$ &
      $-i\psi^b_1$ & $-i\psi^c_1$ & $\psi^c_1$ & $-i\psi^b_1$ 
                 & $\frac{i}{2}$ & $\frac{1}{2} $ & $\frac{i}{2} $ & $\frac{1}{2} $   
                 & $1 $ & $1 $ \\
    $\psi^a_2$ & $[-i] [-]$ & $\psi^a_4$ & $\psi^a_3$ & $-i\psi^a_3$ & $-\psi^a_4$ & 
                       $i\psi^b_2$ & $i\psi^c_2$ & $-\psi^c_2$ & $i\psi^b_2$
                  & $-\frac{i}{2}$ & $-\frac{1}{2} $ & $\frac{i}{2} $ & $\frac{1}{2} $   
                  & $1 $ & $1 $ \\
    $\psi^a_3$ & $[-i] (+)$ & $\psi^a_1$ & $-\psi^a_2$ & $-i\psi^a_2$ & $-\psi^a_1$ & 
      $i\psi^b_3$ & $i\psi^c_3$ & $-\psi^c_3$ & $i\psi^b_3$ 
                    & $-\frac{i}{2}$ & $\frac{1}{2} $ & $\frac{i}{2} $ & $\frac{1}{2} $   
                    & $-1 $ & $1 $ \\
    $\psi^a_4$ & $(+i) [-]$ & $\psi^a_2$ & $-\psi^a_1$ & $i\psi^a_1$ & $\psi^a_2$ &
     $-i\psi^b_4$ & $-i\psi^c_4$ & $\psi^c_4$ & $-i\psi^b_4$ 
                    & $\frac{i}{2}$ & $-\frac{1}{2} $ & $\frac{i}{2} $ & $\frac{1}{2} $   
                    & $-1 $ & $1 $ \\
    \hline
    $\psi^b_1$ & $[+i] (+)$ & $\psi^b_3$ & $-\psi^b_4$ & $-i\psi^b_4$ & $\psi^b_3$ &
     $i\psi^a_1$ & $i\psi^d_1$ & $-\psi^d_1$ & $-i\psi^a_1$ 
                   & $\frac{i}{2}$ & $\frac{1}{2} $ & $-\frac{i}{2} $ & $\frac{1}{2} $   
                   & $1 $ & $-1 $ \\
     $\psi^b_2$ & $(-i) [-]$ & $\psi^b_4$ & $-\psi^b_3$ & $i\psi^b_3$ & $-\psi^b_4$ &
      $-i\psi^a_2$ & $-i\psi^d_2$ & $\psi^d_2$ & $i\psi^a_2$ 
                  & $-\frac{i}{2}$ & $-\frac{1}{2} $ & $-\frac{i}{2} $ & $\frac{1}{2} $   
                  & $1 $ & $-1 $ \\
    $\psi^b_3$ & $(-i) (+)$ & $\psi^b_1$ & $\psi^b_2$ & $i\psi^b_2$ & $-\psi^b_1$ &
      $-i\psi^a_3$ & $-i\psi^d_3$ & $\psi^d_3$ & $i\psi^a_3$ 
                   & $-\frac{i}{2}$ & $\frac{1}{2} $ & $-\frac{i}{2} $ & $\frac{1}{2} $   
                   & $-1 $ & $-1 $ \\
    $\psi^b_4$ & $[+i] [-]$ & $\psi^b_2$ & $\psi^b_1$ & $-i\psi^b_1$ & $\psi^b_2$ &
      $i\psi^a_4$ & $i\psi^d_4$ & $-\psi^d_4$ & $-i\psi^a_4$ 
                   & $\frac{i}{2}$ & $-\frac{1}{2} $ & $-\frac{i}{2} $ & $\frac{1}{2} $   
                   & $-1 $ & $-1 $ \\
    \hline    
    $\psi^c_1$ & $(+i) [+]$ & $\psi^c_3$ & $-\psi^c_4$ & $-i\psi^c_4$ & $\psi^c_3$  &
      $i\psi^d_1$ & $-i\psi^a_1$ & $-\psi^a_1$ & $i\psi^d_1$ 
                  & $\frac{i}{2}$ & $\frac{1}{2} $ & $\frac{i}{2} $ & $-\frac{1}{2} $   
                  & $1 $ & $-1 $ \\
    $\psi^c_2$ & $[-i] (-)$ & $\psi^c_4$ & $-\psi^c_3$ & $i\psi^c_3$ & $-\psi^c_4$& 
      $-i\psi^d_2$ & $i\psi^a_2$ & $\psi^a_2$ & $-i\psi^d_2$ 
                  & $-\frac{i}{2}$ & $-\frac{1}{2} $ & $\frac{i}{2} $ & $-\frac{1}{2} $   
                  & $1 $ & $-1 $ \\
    $\psi^c_3$ & $[-i] [+]$ & $\psi^c_1$ & $\psi^c_2$ & $i\psi^c_2$ & $-\psi^c_1$ &
      $-i\psi^d_3$ & $i\psi^a_3$ & $\psi^a_3$ & $-i\psi^d_3$ 
                  & $-\frac{i}{2}$ & $\frac{1}{2} $ & $\frac{i}{2} $ & $-\frac{1}{2} $   
                  & $-1 $ & $-1 $ \\
    $\psi^c_4$ & $(+i) (-)$ & $\psi^c_{2}$ & $\psi^c_1$ & $-i\psi^c_1$ &  $\psi^c_{2}$ & 
     $i\psi^d_4$ & $-i\psi^a_4$ & $-\psi^a_4$ & $i\psi^d_4$  
                 & $\frac{i}{2}$ & $-\frac{1}{2} $ & $\frac{i}{2} $ & $-\frac{1}{2} $   
                 & $-1 $ & $-1 $ \\
    \hline
    $\psi^d_1$ & $[+i] [+]$ & $\psi^d_3$ & $\psi^d_4$ & $i\psi^d_4$ & $\psi^d_3$& 
       $-i\psi^c_1$ & $i\psi^b_1$ & $\psi^b_1$ & $i\psi^c_1$ 
                & $\frac{i}{2}$ & $\frac{1}{2} $ & $-\frac{i}{2} $ & $-\frac{1}{2} $   
                & $1 $ & $1 $ \\
    $\psi^d_2$ & $(-i) (-)$ & $\psi^d_4$ & $\psi^d_3$ & $-i\psi^d_3$ & $-\psi^d_4$& 
      $i\psi^c_2$ & $-i\psi^b_2$ & $-\psi^b_2$ & $-i\psi^c_2$ 
               & $-\frac{i}{2}$ & $-\frac{1}{2} $ & $-\frac{i}{2} $ & $-\frac{1}{2} $    
               & $1 $ & $1 $ \\
    $\psi^d_3$ & $(-i) [+]$ & $\psi^d_1$ & $-\psi^d_2$ & $-i\psi^d_2$ & $-\psi^d_1$& 
      $i\psi^c_3$ & $-i\psi^b_3$ & $-\psi^b_3$ & $-i\psi^c_3$ 
                & $-\frac{i}{2}$ & $\frac{1}{2} $ & $-\frac{i}{2} $ & $-\frac{1}{2} $    
                & $-1 $ & $1 $ \\
    $\psi^d_4$ & $[+i] (-)$ & $\psi^d_2$ & $-\psi^d_1$ & $i\psi^d_1$ & $\psi^d_2$&
     $-i\psi^c_4$ & $i\psi^b_4$ & $\psi^b_4$ & $i\psi^c_4$ 
                & $\frac{i}{2}$ & $-\frac{1}{2} $ & $-\frac{i}{2} $ & $-\frac{1}{2} $    
                & $-1 $ & $1 $ \\
    \hline 
  \end{tabular}
\end{center}
\caption{\label{tableBasVecd4m} In this table $2^d=16$ vectors, describing internal space 
of fermions in $d=(3+1)$, are presented. Each vector carries the family member quantum 
number  $m=(1,2,3,4)$ --- determined by $S^{03}$ and $S^{12}$, Eqs.~(\ref{cartan}, 
\ref{grapheigen}) --- and the family quantum number $f=(a,b,c,d)$ --- determined by 
$\tilde{S}^{03}$ and $\tilde{S}^{12}$, Eq.~(\ref{cartan}, \ref{grapheigen}). Vectors 
$\psi^f_m$ are obtained by applying $\hat{b}^{f \dagger}_m$ on the vacuum state,
Eq.~(\ref{vac}).
 Vectors, that is the family members of any family, split into even (they are sums of 
products of an even number of $\gamma^a$'s) and odd (they are sums of products of 
an odd number of $\gamma^a$'s). If these vectors are embedded into the vectors 
of $d=(5+1)$ (by being multiplied by an appropriate nilpotent or projector so that they are
of an odd Clifford character), they "gain" charges as presented in Table~\ref{cliff basis5+1.}.} 
\end{tiny}
\end{table}

As seen in Table~\ref{tableBasVecd4m} $\gamma^a$'s  change the handedness 
(Eq.~(\ref{sabgamma})) $\Gamma^{(3+1)}$ of vectors, while  $\tilde{\gamma}^a$'s 
change the handedness $\tilde{\Gamma}^{(3+1)}$ of vectors. Both change the Clifford 
character of states, from Clifford odd character to Clifford even character or vice versa.  
$S^{ab}$, which do not belong to Cartan subalgebra, generate all the states of one 
representation of particular handedness $\Gamma^{(3+1)}$ and particular family 
quantum number. $\tilde{S}^{ab}$, which do not belong to Cartan subalgebra, transform a 
family member of one family into the same family member number of another family, 
$\tilde{\gamma}^{a}$ change the family quantum number as well as the handedness 
$\tilde{\Gamma}^{(3+1)}$.

Dirac  matrices $\gamma^a$ and $S^{ab}$  do not distinguish among the families:
Corresponding family members of any family have the same properties with respect
to $S^{ab}$ and $\gamma^a$, manifesting for $d=(3+1)$ space four times twice 
$2 \times 2$ by diagonal matrices, which are, up to a phase, identical.
The operators  $\gamma^a$ and  $S^{ab}$ are correspondingly four times $4 \times 4$ 
matrices.

One finds among Clifford even vectors of Table~\ref{tableBasVecd4m} the ones which are 
products of projectors; they are Hermitian self conjugated.
The remaining even vectors are  Hermitian conjugated to each other 
($\hat{b}^{a \dagger}_{1}$ is Hermitian conjugated to $\hat{b}^{d \dagger}_{2}$, 
for example). 
In the Clifford odd part of Table~\ref{tableBasVecd4m} one finds that
$\hat{b}^{a \dagger}_{m=(3,4)}$ ($\stackrel{03}{[-i]} \stackrel{12}{(+)}\,, 
\stackrel{03}{(+i)} \stackrel{12}{[-]} $) have as the Hermitian conjugated partners
$\hat{b}^{(c,b) }_{m=2}$ ($-\stackrel{03}{[-i]} \stackrel{12}{(-)}\,$,  
$\stackrel{03}{(-i)}\stackrel{12}{[-]} $), respectively. 
And $\hat{b}^{d \dagger}_{m=(3,4)}$ ($\stackrel{03}{(-i)}\, \stackrel{12}{[+]}\,$, 
$\stackrel{03}{[+i]}\, \stackrel{12}{(-)} $) have as the Hermitian conjugated partners
$\hat{b}^{(c,b)}_{m=1}$ ($\stackrel{03}{(+i)}\, \stackrel{12}{[+]}\,$, 
$-\stackrel{03}{[+i]}\, \stackrel{12}{(+)} $), respectively.

The vacuum state for the $d=(3+1)$  case is correspondingly:\\
$(\frac{1}{\sqrt{2}})^2\,(\stackrel{03}{[-i]} \stackrel{12}{[-]}+ \stackrel{03}{[+i]} 
\stackrel{12}{[+]} + \stackrel{03}{[+i]} \stackrel{12}{[-]} + \stackrel{03}{[-i]} 
\stackrel{12}{[+]})$ (one has to take into account that vectors in $d=(1+3)$ are 
embedded into  $d=(5+1)$). 

Embedding  $\hat{b}^{b \dagger}_{m=3}$ ($=\stackrel{03}{[-i]}\, \stackrel{12}{(+)}$) 
into odd part of Table~\ref{cliff  basis5+1.}, the creation operator extends into 
$\stackrel{03}{[-i]}\, \stackrel{12}{(+)}\,\stackrel{12}{[-]}$, manifesting in $d=(3+1)$ 
the charge $-\frac{1}{2}$ (while the annihilation operator extends into $-\stackrel{03}{[-i]}\,
 \stackrel{12}{(-)}\,\stackrel{12}{[-]}$).

\subsection{$\gamma^a$ matrices in $d=(3+1)$}
\label{basis}
%

There are $2^4 = 16$ basic vectors in  $d=(3+1)$, presented in Table~\ref{tableBasVecd4m}.
They all can be found as well as a part of states in Table~\ref{cliff basis5+1.} with 
either nilpotent or projector, expressing the charge, added so that each state has an odd
Clifford character belonging to one of $16$ vectors of $odd I $ in Table~\ref{cliff basis5+1.}. 
We make a choice of 
products of nilpotents and projectors, which are eigenstates of the Cartan subalgebra 
operators, Eq.~(\ref{cartan}), as presented in Eqs.~(\ref{grapheigen}). 

The family members of a family are reachable by either $S^{ab}$ or by $\gamma^a$, and 
represent twice two vectors of definite handedness $\Gamma^{(d)}$ in $d=(3+1)$. 
Different families are reachable by  either $\tilde{S}^{ab}$ or by $\tilde{\gamma}^a$. 
Each state carries correspondingly quantum numbers of the two kinds of the 
Cartan subalgebra. In Table~\ref{tableBasVecd4m} also 
$\Gamma^{(3+1)}$ ($ = - 4i S^{03} S^{12}$) and 
$\tilde{\Gamma}^{(3+1)}$  ($ = - 4i S^{03} S^{12}$) are presented.
Let us again point out that if we treat all the basic vectors in $d=(3+1)$ as a part of vectors 
in $d=(5+1)$, all of an odd Clifford character, so that they carry also a charge which is  
the spin $S^{56}$, then the family members of a family are reachable by $S^{ab}$
only and families by  $\tilde{S}^{ab}$ only.

When the basic vectors are chosen and Table~\ref{tableBasVecd4m} is made it is not difficult
to find the matrix representations for the operators ($\gamma^a$, $S^{ab}$, $\tilde{\gamma}^{a}$, 
$\tilde{S}^{ab}$, $\Gamma^{(3+1)}$,  $\tilde{\Gamma}^{(3+1)}$). They are obviously 
$16 \times 16$ matrices with a  $4 \times 4$  diagonal or  off diagonal 
 or partly diagonal and partly off diagonal substructure.

 %
Let us define, to simplify the notation, the unit $4 \times 4$ submatrix and the submatrix 
with all the matrix elements equal to zero as follows
 \begin{equation}\label{dn-mat-mat1and0}
     \mathbf{1}= 
   \begin{pmatrix}
       1 & 0 \\
       0 & 1 
    \end{pmatrix}  \, = \sigma^0,
\qquad\qquad
%
  %
    \mathbf{0}= 
   \begin{pmatrix}
       0 & 0 \\
       0 & 0 
    \end{pmatrix}.
  \end{equation}
 We also use ($2\times 2$) Pauli matrices
 \begin{equation}\label{pauli}
   \sigma^1=
   \begin{pmatrix}
     0 & 1\\
     1 & 0
    \end{pmatrix}, \qquad
   \sigma^2=
   \begin{pmatrix}
     0 & -i\\
     i & 0
    \end{pmatrix}, \qquad
   \sigma^3=
   \begin{pmatrix}
     1 & 0\\
     0 & -1
    \end{pmatrix}.
 \end{equation}

It is easy to find the matrix representations for $\gamma^0$, $\gamma^1$, 
$\gamma^2$ and $\gamma^3$ from Table~\ref{tableBasVecd4m} 
 \begin{equation}\label{dn-mat-mgamma01}
 \gamma^{0} =
 \begin{pmatrix}
      \begin{smallmatrix}
          0 & \sigma^0\\
          \sigma^0 & 0 
        \end{smallmatrix} & \mathbf{0}  & \mathbf{0}  & \mathbf{0} \\
        \mathbf{0}  &
        \begin{smallmatrix}
          0 & \sigma^0\\
          \sigma^0 & 0 
        \end{smallmatrix} & \mathbf{0}  & \mathbf{0} \\
        \mathbf{0} & \mathbf{0}  &
        \begin{smallmatrix}
          0 & \sigma^0\\
          \sigma^0 & 0 
        \end{smallmatrix} & \mathbf{0} \\
        \mathbf{0} & \mathbf{0} & \mathbf{0} &
        \begin{smallmatrix}
          0 & \sigma^0\\
          \sigma^0 & 0 
        \end{smallmatrix}
  \end{pmatrix}\,, 
%
%
 \,\gamma^{1} =
 \begin{pmatrix}
      \begin{smallmatrix}
          0 & \sigma^1\\
          -\sigma^1 & 0
        \end{smallmatrix} & \mathbf{0} & \mathbf{0} & \mathbf{0}\\
        \mathbf{0} &
        \begin{smallmatrix}
          0 & -\sigma^1\\
          \sigma^1 & 0
        \end{smallmatrix} & \mathbf{0} & \mathbf{0}\\
        \mathbf{0} & \mathbf{0} &
        \begin{smallmatrix}
          0 & -\sigma^1\\
          \sigma^1 & 0
        \end{smallmatrix} & \mathbf{0} \\
        \mathbf{0} & \mathbf{0} & \mathbf{0} &
        \begin{smallmatrix}
          0 & \sigma^1\\
          -\sigma^1 & 0
        \end{smallmatrix}
  \end{pmatrix}\,, \nonumber\\
 \end{equation}
 \begin{equation}\label{dn-mat-mgamma23}
 \gamma^{2} =
 \begin{pmatrix}
      \begin{smallmatrix}
          0 & -\sigma^2\\
          \sigma^2 & 0 & 
        \end{smallmatrix} & 0 & 0 & 0\\
        \mathbf{0} &
        \begin{smallmatrix}
          0 & \sigma^2\\
          -\sigma^2 & 0 & 
        \end{smallmatrix} & \mathbf{0} & \mathbf{0}\\
        \mathbf{0} & \mathbf{0} &
        \begin{smallmatrix}
          0 & \sigma^2\\
          -\sigma^2 & 0 & 
        \end{smallmatrix} & \mathbf{0} \\
        \mathbf{0} & \mathbf{0} & \mathbf{0} &
        \begin{smallmatrix}
          0 & -\sigma^2\\
          \sigma^2 & 0 & 
        \end{smallmatrix}
  \end{pmatrix}\,,
%
 %
\,
 \gamma^{3} =
 \begin{pmatrix}
      \begin{smallmatrix}
          0 & \sigma^3\\
          -\sigma^3 & 0 & 
        \end{smallmatrix} & \mathbf{0} & \mathbf{0} & \mathbf{0}\\
        \mathbf{0} &
        \begin{smallmatrix}
          0 & \sigma^3\\
          -\sigma^3 & 0 & 
        \end{smallmatrix} & \mathbf{0} & \mathbf{0}\\
        \mathbf{0} & \mathbf{0} &
        \begin{smallmatrix}
          0 & \sigma^3\\
          -\sigma^3 & 0 & 
        \end{smallmatrix} & \mathbf{0} \\
        \mathbf{0} & \mathbf{0} & \mathbf{0} &
        \begin{smallmatrix}
          0 & \sigma^3\\
          -\sigma^3 & 0 & 
        \end{smallmatrix}
  \end{pmatrix}\,, 
 \end{equation}
manifesting  the $4 \times 4$ substructure along the diagonal of $16 \times 16$
matrices.

The representations of the $\tilde{\gamma}^{a}$ do not appear in the Dirac case.
They manifest the off diagonal structure as follows
 \begin{equation}\label{dn-mat-mtgamma01}
 \Tilde{\gamma}^{0} =
 \begin{pmatrix}
      \mathbf{0} & 
      \begin{smallmatrix}
          -i\sigma^3 & 0 \\
          0 & i\sigma^3
        \end{smallmatrix} & \mathbf{0} & \mathbf{0} \\
        \begin{smallmatrix}
          i\sigma^3 & 0 \\
          0 & -i\sigma^3
        \end{smallmatrix} & \mathbf{0} & \mathbf{0} & \mathbf{0}\\
        \mathbf{0} & \mathbf{0} &  \mathbf{0} & 
        \begin{smallmatrix}
          i\sigma^3 & 0 \\
          0 & -i\sigma^3
        \end{smallmatrix}  \\
        \mathbf{0} & \mathbf{0}  &
        \begin{smallmatrix}
          -i\sigma^3 & 0 \\
          0 & i\sigma^3
        \end{smallmatrix} & \mathbf{0} 
  \end{pmatrix}
\,,
%
%
\, \Tilde{\gamma}^{1} =
 \begin{pmatrix}
        \mathbf{0} & \mathbf{0} &
        \begin{smallmatrix}
          -i\sigma^3 & 0 \\
          0 & i\sigma^3
        \end{smallmatrix} & \mathbf{0} \\
        \mathbf{0} & \mathbf{0} & \mathbf{0} & 
        \begin{smallmatrix}
          i\sigma^3 & 0 \\
          0 & -i\sigma^3
        \end{smallmatrix} \\
        \begin{smallmatrix}
          -i\sigma^3 & 0 \\
          0 & i\sigma^3
        \end{smallmatrix} & \mathbf{0}  & \mathbf{0} & \mathbf{0} \\
        \mathbf{0} & 
        \begin{smallmatrix}
          i\sigma^3 & 0 \\
          0 & -i\sigma^3
        \end{smallmatrix} & \mathbf{0} & \mathbf{0}
  \end{pmatrix}\,, \nonumber\\
 \end{equation}
 \begin{equation}\label{dn-mat-mtgamma23}
 \Tilde{\gamma}^{2} =
 \begin{pmatrix}
       \mathbf{0} & \mathbf{0} &
      \begin{smallmatrix}
           \sigma^3 & 0 \\
          0 & -\sigma^3
       \end{smallmatrix} & \mathbf{0} \\
        \mathbf{0} & \mathbf{0} & \mathbf{0} &
        \begin{smallmatrix}
          -\sigma^3 & 0 \\
          0 & \sigma^3
        \end{smallmatrix} \\
        \begin{smallmatrix}
          -\sigma^3 & 0 \\
          0 & \sigma^3
        \end{smallmatrix} & \mathbf{0}  & \mathbf{0} & \mathbf{0} \\
        \mathbf{0} & 
        \begin{smallmatrix}
          \sigma^3 & 0 \\
          0 & -\sigma^3
        \end{smallmatrix} & \mathbf{0} & \mathbf{0}
  \end{pmatrix}\,, 
%
 %
\, \Tilde{\gamma}^{3} =
 \begin{pmatrix}
        \mathbf{0} & 
       \begin{smallmatrix}
          -i\sigma^3 & 0 \\
          0 & i\sigma^3
        \end{smallmatrix} & \mathbf{0} & \mathbf{0} \\
        \begin{smallmatrix}
          -i\sigma^3 & 0 \\
          0 & i\sigma^3
        \end{smallmatrix} & \mathbf{0} & \mathbf{0} & \mathbf{0}\\
        \mathbf{0} & \mathbf{0} &  \mathbf{0} & 
        \begin{smallmatrix}
          -i\sigma^3 & 0 \\
          0 & i\sigma^3
        \end{smallmatrix}\\
        \mathbf{0} & \mathbf{0} & 
        \begin{smallmatrix}
          -i\sigma^3 & 0 \\
          0 & i\sigma^3
        \end{smallmatrix} & \mathbf{0}
  \end{pmatrix}\,.
 \end{equation}
%

Matrices $S^{ab}$ have again along the diagonal the $4 \times 4$ substructure, 
as expected. They manifest the repetition of the Dirac $4 \times 4$ matrices, up to a phase,
since the Dirac $S^{ab}$ do not distinguish among families.
 \begin{equation}\label{dn-mat-mS0102}
 S^{01} = \frac{i}{2}
 \begin{pmatrix}
      \begin{smallmatrix}
          \sigma^1 & 0 \\
          0 & -\sigma^1
        \end{smallmatrix} & \mathbf{0} & \mathbf{0} & \mathbf{0}\\
        \mathbf{0} &
        \begin{smallmatrix}
          -\sigma^1 & 0 \\
          0 & \sigma^1
        \end{smallmatrix} & \mathbf{0} & \mathbf{0}\\
        \mathbf{0} & \mathbf{0} &
        \begin{smallmatrix}
          -\sigma^1 & 0 \\
          0 & \sigma^1
        \end{smallmatrix} & \mathbf{0} \\
        \mathbf{0} & \mathbf{0} & \mathbf{0} &
        \begin{smallmatrix}
          \sigma^1 & 0 \\
          0 & -\sigma^1
        \end{smallmatrix}
  \end{pmatrix}\,, 
%
%
\, S^{02} = \frac{i}{2}
 \begin{pmatrix}
      \begin{smallmatrix}
          -\sigma^2 & 0 \\
          0 & \sigma^2
        \end{smallmatrix} & \mathbf{0} & \mathbf{0} & \mathbf{0}\\
        \mathbf{0} &
        \begin{smallmatrix}
          \sigma^2 & 0 \\
          0 & -\sigma^2
        \end{smallmatrix} & \mathbf{0} & \mathbf{0}\\
        \mathbf{0} & \mathbf{0} &
        \begin{smallmatrix}
          \sigma^2 & 0 \\
          0 & -\sigma^2
        \end{smallmatrix} & \mathbf{0} \\
        \mathbf{0} & \mathbf{0} & \mathbf{0} &
        \begin{smallmatrix}
          -\sigma^2 & 0 \\
          0 & \sigma^2
        \end{smallmatrix}
  \end{pmatrix}\,,  \nonumber\\
 \end{equation}
 \begin{equation}\label{dn-mat-mS0312}
 S^{03} = \frac{i}{2}
 \begin{pmatrix}
      \begin{smallmatrix}
          \sigma^3 & 0 \\
          0 & -\sigma^3
        \end{smallmatrix} & \mathbf{0} & \mathbf{0} & \mathbf{0}\\
        \mathbf{0} &
        \begin{smallmatrix}
          \sigma^3 & 0 \\
          0 & -\sigma^3
        \end{smallmatrix} & \mathbf{0} & \mathbf{0}\\
        \mathbf{0} & \mathbf{0} &
        \begin{smallmatrix}
          \sigma^3 & 0 \\
          0 & -\sigma^3
        \end{smallmatrix} & \mathbf{0} \\
        \mathbf{0} & \mathbf{0} & \mathbf{0} &
        \begin{smallmatrix}
          \sigma^3 & 0 \\
          0 & -\sigma^3
        \end{smallmatrix}
  \end{pmatrix}\,,  
%
%
\, S^{12} = \frac{1}{2}
 \begin{pmatrix}
      \begin{smallmatrix}
          \sigma^3 & 0 \\
          0 & \sigma^3
        \end{smallmatrix} & \mathbf{0} & \mathbf{0} & \mathbf{0}\\
        \mathbf{0} &
        \begin{smallmatrix}
          \sigma^3 & 0 \\
          0 & \sigma^3
        \end{smallmatrix} & \mathbf{0} & \mathbf{0}\\
        \mathbf{0} & \mathbf{0} &
        \begin{smallmatrix}
          \sigma^3 & 0 \\
          0 & \sigma^3
        \end{smallmatrix} & \mathbf{0} \\
        \mathbf{0} & \mathbf{0} & \mathbf{0} &
        \begin{smallmatrix}
          \sigma^3 & 0 \\
          0 & \sigma^3
        \end{smallmatrix}
  \end{pmatrix}\,, \nonumber\\
 \end{equation}
 \begin{equation}\label{dn-mat-mS1323}
 S^{13} = \frac{1}{2}
 \begin{pmatrix}
      \begin{smallmatrix}
          \sigma^2 & 0 \\
          0 & \sigma^2
        \end{smallmatrix} & \mathbf{0} & \mathbf{0} & \mathbf{0}\\
        \mathbf{0} &
        \begin{smallmatrix}
          -\sigma^2 & 0 \\
          0 & -\sigma^2
        \end{smallmatrix} & \mathbf{0} & \mathbf{0}\\
        \mathbf{0} & \mathbf{0} &
        \begin{smallmatrix}
          -\sigma^2 & 0 \\
          0 & -\sigma^2
        \end{smallmatrix} & \mathbf{0} \\
        \mathbf{0} & \mathbf{0} & \mathbf{0} &
        \begin{smallmatrix}
          \sigma^2 & 0 \\
          0 & \sigma^2
        \end{smallmatrix}
  \end{pmatrix}\,, 
%
%
 S^{23} = \frac{1}{2} 
 \begin{pmatrix}
      \begin{smallmatrix}
          \sigma^1 & 0 \\
          0 & \sigma^1
        \end{smallmatrix} & \mathbf{0} & \mathbf{0} & \mathbf{0}\\
        \mathbf{0} &
        \begin{smallmatrix}
          -\sigma^1 & 0 \\
          0 & -\sigma^1
        \end{smallmatrix} & \mathbf{0} & \mathbf{0}\\
        \mathbf{0} & \mathbf{0} &
        \begin{smallmatrix}
          -\sigma^1 & 0 \\
          0 & -\sigma^1
        \end{smallmatrix} & \mathbf{0} \\
        \mathbf{0} & \mathbf{0} & \mathbf{0} &
        \begin{smallmatrix}
          \sigma^1 & 0 \\
          0 & \sigma^1
        \end{smallmatrix}
      \end{pmatrix}\,.
\end{equation}
\begin{equation}\label{dn-mat-mGamma31}
  \Gamma^{(3+1)}=-4i S^{03}S^{12}=
 \begin{pmatrix}
      \begin{smallmatrix}
          1 & 0 \\
          0 & -1 
        \end{smallmatrix} & \mathbf{0} & \mathbf{0} & \mathbf{0}\\
        \mathbf{0} &
        \begin{smallmatrix}
          1 & 0 \\
          0 & -1 
        \end{smallmatrix} & \mathbf{0} & \mathbf{0}\\
        \mathbf{0} & \mathbf{0} &
        \begin{smallmatrix}
          1 & 0 \\
          0 & -1 
        \end{smallmatrix} & \mathbf{0} \\
        \mathbf{0} & \mathbf{0} & \mathbf{0} &
        \begin{smallmatrix}
          1 & 0 \\
          0 & -1 
        \end{smallmatrix}
  \end{pmatrix}\,.
\end{equation}
 
The operators $\tilde{S}^{ab}$ have again off diagonal $4 \times 4$ substructure, except
$\tilde{S}^{03}$ and $\tilde{S}^{12}$, which are diagonal.
 \begin{equation}\label{dn-mat-mtS010203}
 \Tilde{S}^{01} = -\frac{i}{2}
 \begin{pmatrix}
      \mathbf{0} & \mathbf{0} & \mathbf{0} &
      \mathbf{1}\\
        \mathbf{0} & \mathbf{0} & 
        \mathbf{1} & \mathbf{0} \\
        \mathbf{0} & 
        \mathbf{1} & \mathbf{0}  & \mathbf{0} \\
        \mathbf{1} & \mathbf{0} & \mathbf{0} & \mathbf{0}
  \end{pmatrix}\,, 
%
 %
\, \Tilde{S}^{02} = \frac{1}{2} 
  \begin{pmatrix}
      \mathbf{0} & \mathbf{0} & \mathbf{0} &
       \mathbf{1}\\
        \mathbf{0} & \mathbf{0} & 
        \mathbf{1} & \mathbf{0} \\
        \mathbf{0} & 
        -\mathbf{1} & \mathbf{0}  & \mathbf{0} \\
        -\mathbf{1} & \mathbf{0} & \mathbf{0} & \mathbf{0}
  \end{pmatrix}\,,  
%
 %
\, \Tilde{S}^{03} = \frac{i}{2}
 \begin{pmatrix}
      \mathbf{1} & \mathbf{0} & \mathbf{0} & \mathbf{0}\\
        \mathbf{0} &
        -\mathbf{1} & \mathbf{0} & \mathbf{0}\\
        \mathbf{0} & \mathbf{0} &
        \mathbf{1} & \mathbf{0} \\
        \mathbf{0} & \mathbf{0} & \mathbf{0} &
        -\mathbf{1}
  \end{pmatrix}\,,  
\end{equation}
\begin{equation}\label{dn-mat-mtS121323}
\, \Tilde{S}^{12} = \frac{1}{2} 
 \begin{pmatrix}
       \mathbf{1} & \mathbf{0} & \mathbf{0} & \mathbf{0}\\
        \mathbf{0} &
        \mathbf{1}  & \mathbf{0} & \mathbf{0}\\
        \mathbf{0} & \mathbf{0} &
        -\mathbf{1}  & \mathbf{0} \\
        \mathbf{0} & \mathbf{0} & \mathbf{0} &
        -\mathbf{1} 
      \end{pmatrix}\,,  
%
 %
\, \Tilde{S}^{13} = \frac{i}{2}
   \begin{pmatrix}
      \mathbf{0} & \mathbf{0} & \mathbf{0} &
      -\mathbf{1}  \\
        \mathbf{0} & \mathbf{0} & 
        \mathbf{1} & \mathbf{0} \\
        \mathbf{0} & 
        -\mathbf{1} & \mathbf{0}  & \mathbf{0} \\
        \mathbf{1} & \mathbf{0} & \mathbf{0} & \mathbf{0}
  \end{pmatrix}\,, 
%
 %
\, \Tilde{S}^{23} = \frac{1}{2}
   \begin{pmatrix}
      \mathbf{0} & \mathbf{0} & \mathbf{0} &
      -\mathbf{1}  \\
        \mathbf{0} & \mathbf{0} & 
        \mathbf{1} & \mathbf{0} \\
        \mathbf{0} & 
        \mathbf{1} & \mathbf{0}  & \mathbf{0} \\
        -\mathbf{1} & \mathbf{0} & \mathbf{0} & \mathbf{0}
  \end{pmatrix}\,.
 \end{equation}

 \begin{equation}\label{dn-mat-mtGamma31}
   \Tilde{\Gamma}^{(3+1)}=-4i \Tilde{S}^{03}\Tilde{S}^{12}=
 \begin{pmatrix}
        \mathbf{1} & \mathbf{0} & \mathbf{0} & \mathbf{0}\\
        \mathbf{0} &
        -\mathbf{1} & \mathbf{0} & \mathbf{0}\\
        \mathbf{0} & \mathbf{0} &
        -\mathbf{1} & \mathbf{0} \\
        \mathbf{0} & \mathbf{0} & \mathbf{0} &
        \mathbf{1}
  \end{pmatrix}\,.
\end{equation}

\section{Conclusions}
\label{conclusions}
%

We present in this contribution the matrix representations of operators, $\gamma^a$'s, 
$S^{ab}$'s, $\tilde{\gamma}^a$'s, $\tilde{S}^{ab}$'s, applying on
a basis, and defined by the creation and annihilation operators for fermions in $d$-dimensional
Clifford space, where  $d=2(2n+1)$, or $4n$, $n=1$.  We make a choice of 
$d=(3+1)$ and $d=(5+1)$, paying attention that creation and annihilation operators of
fermions in $d=(3+1)$ manifest charges, if embedded in $d=(5+1)$.

Creation and annihilation operators, defining the internal space of the second quantized 
fermions, have an odd Clifford character (they are superposition of
odd products of Clifford objects $\gamma^a$'s). They are in our presentation products 
of nilpotents and projectors, chosen to be eigenvectors of  the Cartan subalgebra, 
Eq.~(\ref{cartan}), of the Lorentz algebra of $S^{ab}$, as well as of the corresponding 
Cartan subalgebra, Eq.~(\ref{cartan}), of  $\tilde{S}^{ab}$.  $S^{ab}$ define the
members of each irreducible representation of the Lorentz group, $\tilde{S}^{ab}$ define
family quantum number of each irreducible representation.  

Creation and annihilation operators are Hermitian conjugated to each other. We 
make a choice of the creation operators with respect to the annihilation operators by 
choosing the vacuum state, Eq.~(\ref{vac}),
to be  the sum of products of the annihilation operators with their Hermitian 
conjugated partners creation operators. 

$S^{ab}$ generate $2^{\frac{d}{2}-1}$  family members of a particular family of an odd 
Clifford character, $\tilde{S}^{ab}$ generate the corresponding $2^{\frac{d}{2}-1}$ families. 
The Hermitian conjugation determines their $2^{\frac{d}{2}-1}\times$ $2^{\frac{d}{2}-1}$
partners (which are reachable also by $\gamma^a \tilde{\gamma}^a$). The Clifford even
representations follow from the odd $2^{d-1}$ vectors by the application of  $\gamma^a$'s 
or $ \tilde{\gamma}^a$'s.
There are correspondingly $2^d$ vectors in $d$-dimensional space ($d=2(2n+1),4n$). 

The Clifford even operators, $S^{ab}$ and $\tilde{S}^{ab}$, keep the Clifford character 
unchanged.  
$\gamma^a$'s and $\tilde{\gamma}^a$'s change the Clifford character of  vectors --- from 
odd to even or vice versa.

Embedding $SO(3+1)$ into $SO(d)$, $d > (3+1)$, $d$ even,  makes spins in $d \ge (5+1)$
to manifest in $d=(3+1)$ as charges.

One can check that the creation operators of an odd Clifford 
character and their Hermitian conjugated partners, applied on the vacuum state, 
Eq.(\ref{vac}), fulfill the anticommutation relations for the second quantized fermions, 
Eq.~(\ref{creationvac}), postulated by Dirac, what explains the Dirac's second 
quantization postulates. 

One can also observe the appearance of families, used in the {\it spin-charge-family} theory
to explain families of quarks and leptons~\cite{n2014matterantimatter,IARD2016,
normaJMP2015}, when the Clifford space in $d=(3+1)$ is embedded into $d=(13 +1)$.

There are $2^4 = 16$ basic vectors in  $d=(3+1)$  and correspondingly all the matrices 
have dimension $16 \times 16$, which are for the operators, determined by $\gamma^a$'s,
by diagonal and for the operators, determined by $\tilde{\gamma}^a$'s, off diagonal, except
$\tilde{S}^{03}$, $\tilde{S}^{12}$, which are the members 
of the Cartan subalgebra and correspondingly also $\tilde{\Gamma}^{(3+1)}=-4i \tilde{S}^{03}
\tilde{S}^{12}$. 
We keep the Clifford odd and the Clifford even vectors as the basic vectors. We treat in the 
Clifford odd part the creation and annihilation operators as they would all define the vector 
space, to point out, that if space of $d=(3+1)$ is embedded in $d\ge 6$,
 all the parts, even and odd, contribute to the enlarged vector space as factors.

\begin{acknowledgements}
The author N.S.M.B. thanks Department of Physics, FMF, University of Ljubljana, Society of 
Mathematicians, Physicists and Astronomers of Slovenia,  for supporting the research on the 
{\it spin-charge-family} theory, and  Matja\v z Breskvar of 
Beyond Semiconductor for donations, in particular for sponsoring the annual workshops entitled 
"What comes beyond the standard models" at Bled. 
\end{acknowledgements}


\end{document}